\documentclass[intlimits,twoside,a4paper]{article}
\usepackage{wrapfig}
\usepackage{graphicx}
\usepackage{amsmath,amssymb,bm}
\usepackage[T2A]{fontenc}
\usepackage[cp1251]{inputenc}

\usepackage[eqsecnum]{cmpj3}

\DeclareMathOperator{\Sp}{Sp}
\newcommand{\be}{\begin{equation}}
\newcommand{\ee}{\end{equation}}
\newcommand{\bea}{\begin{eqnarray}}
\newcommand{\eea}{\end{eqnarray}}
\newcommand{\non}{\nonumber\\}

\issue{2017}{20}{4}{43707}
\doinumber{10.5488/CMP.20.43707}

\title[The effect of hydrostatic pressure]{The effect of hydrostatic pressure  on  thermodynamic characteristics of NH$_3$CH$_2$COOH$\cdot$H$_2$PO$_3$ type ferroelectric materials}
\author[I.R. Zachek, R.R. Levitskii, A.S. Vdovych]{ I.R. Zachek\refaddr{label1}, R.R. Levitskii\refaddr{label2}, A.S. Vdovych\refaddr{label2}
}
\addresses{\addr{label1} Lviv Polytechnic National
University, 12 Bandera St., 79013 Lviv, Ukraine
\addr{label2} Institute for Condensed Matter Physics of the
National Academy of Sciences of Ukraine,\\ 1 Svientsitskii St.,
79011 Lviv, Ukraine }

\date{Received June 30, 2017, in final form August 30, 2017}

\begin{document}

\maketitle
\vspace{-1mm}
\begin{abstract}

The model of NH$_3$CH$_2$COOH$\cdot$H$_2$PO$_3$, modified by taking into account the piezoelectric coupling between the ordering structure elements and the strains $\varepsilon_i$, $\varepsilon_j$, is used for investigation of the effects that appear under external pressures.
Within two-particle cluster approximation,
the components of polarization vector and static dielectric permittivity tensor of the mechanically clamped and free crystals, their piezoelectric and thermal characteristics are calculated. The
effect of hydrostatic pressure on the phase transition
and the calculated physical characteristics of the crystal is studied. A good quantitative
description of  experimental data for these crystals is obtained.

\keywords ferroelectrics, phase transition, dielectric permittivity, piezoelectric coefficients, hydrostatic pressures
\pacs 77.22.-d, 77.22.Ch, 77.22.Ej, 77.65.-j, 77.80.Bh
\end{abstract}

\vspace{-4mm}
\section{Introduction}

The study of the effects that appear under external pressures is one of the urgent problems in the physics of ferroelectric materials.
External pressures can be a powerful tool for a purposeful influence on their physical characteristics and can be used in technological processes. The study of the behaviour of ferroelectrics under external pressures enabled us to better understand the mechanisms of phase transitions in these materials.

It is necessary to note that an acceptable description of an external hydrostatic pressure effect on the  phase transition and physical characteristics for many ferroelectric crystals of  KH$_2$PO$_4$ family was made in \cite{kn2009,lev2}, for quasione-dimentional CsH$_2$PO$_4$ type ferroelectrics --- in   \cite{lev3}, for monoclinic RbD$_2$PO$_4$ --- in \cite{zac}, for  RbH$_2$SO$_4$ crystal --- in \cite{zac2}.

In \cite{Zachek_PB2017}, on the basis of the proposed model of deformed NH$_3$CH$_2$COOH$\cdot$H$_2$PO$_3$ (GPI) type ferroelectrics, the dielectric, piezoelectric, elastic and thermal  characteristics of these crystals
 were calculated in the two-particle cluster approximation and a good quantitative description of the available experimental data for these characteristics was obtained. The effect of electric fields on dielectric properties of GPI ferroelectric was investigated in \cite{Zachek_CMP2017}. A satisfactory quantitative description of the corresponding experimental data  was obtained at the proper choice of the model parameters. An experimental study of hydrostatic pressure effect  on the physical properties of GPI type crystals was carried out in \cite{yas,yas1}.
Calculation of the static dielectric permittivities and investigation of the electric field  $E_3$ on the permittivity $\varepsilon_{33}$ of GPI crystal was carried out in \cite{sta1,bal} within the phenomenological Landau theory.

In the present work, the hydrostatic pressure effect on the  phase transition, thermodynamic, dielectric, piezoelectric and elastic characteristics of this type of crystals is studied based on the model of a deformed GPI crystal \cite{Zachek_PB2017}.

\section{ Model Hamiltonian}

\begin{figure}[!b]
\begin{center}
\includegraphics[scale=0.5]{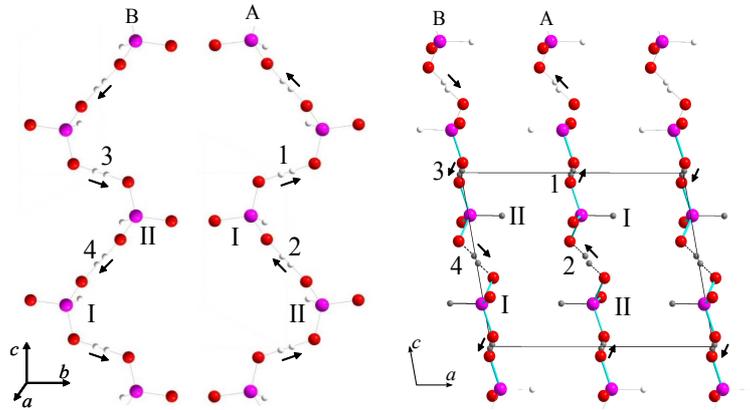}
\end{center}
\caption{(Color online) Orientations of vectors ${\bf d}_{qf}$ in the primitive cell
in the ferroelectric phase \cite{Zachek_PB2017,Zachek_CMP2017}.} \label{struktura}
\end{figure}

We consider the system of protons in GPI, localized on O-H{\ldots}O bonds between phosphite groups HPO$_{3}$, which form zigzag chains along the $c$-axis of the crystal \cite{Zachek_PB2017,Zachek_CMP2017} (figure~\ref{struktura}).
Dipole moments ${ \bf{d}}_{qf}$ (\textit{q} is the number of a primitive cell, $f=1,\dots,4)$ are ascribed to the protons on the bonds. In the ferroelectric phase, the dipole moments compensate each other  (${\bf d}_{q1}$ with ${\bf d}_{q3}$, ${\bf d}_{q2}$ with ${\bf d}_{q4}$) in the directions  $Z$ and $X$ ($X\perp (b,c)$, $Y \parallel b$, $Z \parallel c$), and simultaneously supplement each other in the direction $Y$, creating a spontaneous polarization. Vectors ${\bf d}_{qf}$ are oriented at some angles to crystallographic axes and have longitudinal and transverse components along the $b$-axis.
Herein below, for components of vectors and tensors we often use the notations $1$, $2$ and $3$ instead of $x$, $y$ and $z$ for convenience.
The Hamiltonian of proton subsystem of GPI, which takes into account the short-range and long-range interactions, applied hydrostatic pressure $p=-\sigma_i$ ($i=1, 2, 3$) and electric fields $E_1$, $E_2$, $E_3$ along positive directions of the Cartesian axes $OX$, $OY$ and $OZ$, consists of ``seed'' and pseudospin parts. The ``seed'' energy $U_{\text{seed}}$ corresponds to the heavy ion sublattice and does not depend explicitly on the configuration of the proton subsystem. The pseudospin part takes into account the  short-range $\hat H_{\text{short}}$ and long-range $\hat H_{\text{MF}}$ interactions of protons near tetrahedra  HPO$_3$, as well as the effective interaction with the electric fields  $E_1$, $E_2$ and $E_3$. Therefore \cite{Zachek_PB2017},
\bea
&& \hat H= N U_{\text{ seed}} + \hat H_{\text{ short}} + \hat H_{\text{MF}}\,,
\eea
where $N$  is the total number of primitive cells.

The $U_{ \text{seed}}$ corresponds to the ``seed'' energy, which includes the elastic, piezoelectric and dielectric parts, expressed in terms of electric fields  $E_i$ $(i=1, 2, 3)$ and strains  $\varepsilon_i$ and $\varepsilon_j$
$(j=i+3)$:
\begin{align}
U_{\text{seed}}&= v\Bigg[\frac{1}{2}\sum\limits_{i,i'=1}^3c_{ii'}^{E0}(T)\varepsilon_i \varepsilon_{i'}+
\frac{1}{2}\sum\limits_{j=4}^6c_{jj}^{E0}(T)\varepsilon_j^{2}  + \sum\limits_{i=1}^3c_{i5}^{E0}(T)\varepsilon_i\varepsilon_5 + c_{46}^{E0}(T)\varepsilon_4\varepsilon_6  \nonumber\\
 &\quad -\sum\limits_{i=1}^3 e_{2i}^0 \varepsilon_i E_2 - e_{25}^0 \varepsilon_5 E_2  - e_{14}^0 \varepsilon_4 E_1 - e_{16}^0 \varepsilon_6 E_1-  e_{34}^0 \varepsilon_4 E_3 - e_{36}^0 \varepsilon_6 E_3 
  \nonumber\\
& \quad- \frac{1}{2}  \chi_{11}^{\varepsilon 0}E_1^2 - \frac{1}{2}
\chi_{22}^{\varepsilon 0}E_2^2 -
\frac{1}{2}  \chi_{33}^{\varepsilon 0}E_3^2 - \chi_{31}^{\varepsilon 0}E_3E_1\Bigg]. 
\end{align}
Here, parameters $c_{ii'}^{E0}(T)$, $c_{i5}^{E0}(T)$, $c_{46}^{E0}(T)$, $c_{jj}^{E0}(T)$, $e_{ii'}^0$, $e_{ij}^0$, $\chi_{ii}^{\varepsilon 0}$, $\chi_{31}^{\varepsilon 0}$ $(i'=1, 2, 3)$ correspond to the so-called ``seed'' elastic constants,
piezoelectric stresses and dielectric susceptibilities, respectively, $v$ is the volume of a primitive cell.

The Hamiltonian of short-range interactions
\be
\hat H_{ \text{short}} = -  2w  \sum\limits_{qq'}  \left ( \frac{\sigma_{q1}}{2} \frac{\sigma_{q2}}{2}  +  \frac{\sigma_{q3}}{2}\frac{\sigma_{q4}}{2} \right)
 \bigl( \delta_{{\bf R}_q{\bf R}_{q'}}  +  \delta_{{\bf R}_q + {\bf R}_{c},{\bf R}_{q'}}  \bigr). \label{Hshort}
\ee
In (\ref{Hshort})  $\sigma_{qf}$ is the \textit{z}-component of pseudospin operator that describes the state of the $f$-th bond ($f = 1, 2, 3, 4$), in the  $q$-th cell.
The first Kronecker delta corresponds to the interaction between protons in the chains near the tetrahedra HPO$_{3}$ of type ``I'', while the second  one --- near the tetrahedra HPO$_{3}$ of type ``II'', ${\bf R}_{c}$ is the lattice vector along $c$-axis. The contributions into the energy of interactions between protons near tetrahedra of different type, as well as the mean values of the pseudospins  $\eta_f=\langle \sigma_{qf}\rangle$, which are related to tetrahedra of different type, are equal.

Parameter $w$, which describes the short-range interactions within the chains, is expanded linearly into a series over strains $\varepsilon_i$, $\varepsilon_j$:
\be
w = w^{0} + \sum\limits_{i=1}^3 \delta_{i}\varepsilon_i + \sum\limits_{j=4}^6 \delta_{j}\varepsilon_i. \label{w}
\ee

Hamiltonian $\hat H_{\text{MF}}$ of the long-range dipole-dipole interactions and indirect  (through the lattice vibrations)  interactions between protons in the mean field approximation
takes into account that Fourier transforms of interaction constants $J_{ff'} = \sum\nolimits_{q'} J_{ff'}(qq')$ at ${\bf k} =0$ are linearly expanded over the strains $\varepsilon_i, \varepsilon_j$;
and can be written as:
\be \hat H_{\text{MF}} = N H^{0} + \hat H_s\,,
 \ee where
\begin{align}   H^{0}  &=
   \frac18 J^{0}_{11}(\eta_1^2 + \eta_3^2) +\frac18 J^{0}_{22}(\eta_2^2  + \eta_4^2) + \frac14 J_{13}^{0}\eta_1\eta_3 
 +\frac14 J_{24}^{0}\eta_2\eta_4 +\frac14 J_{12}^{0}(\eta_1\eta_2 + \eta_3\eta_4) +
\frac14 J_{14}^{0}(\eta_1\eta_4 + \eta_2\eta_3) \nonumber \\
& \quad+ \frac18 \Bigg(\sum\limits_{i=1}^3\psi_{11i}\varepsilon_i+ \sum\limits_{j=4}^6\psi_{11j}\varepsilon_j\Bigg)(\eta_1^2 + \eta_3^2) + \frac18 \Bigg(\sum\limits_{i=1}^3\psi_{22i}\varepsilon_i+ \sum\limits_{j=4}^6\psi_{22j}\varepsilon_j\Bigg)(\eta_2^2  + \eta_4^2) \nonumber\\
&\quad + \frac14 \Bigg(\sum\limits_{i=1}^3
 \psi_{13i}\varepsilon_i + \sum\limits_{j=4}^6\psi_{13j}\varepsilon_j\Bigg)\eta_1\eta_3 +\frac14 \Bigg(\sum\limits_{i=1}^3
 \psi_{24i}\varepsilon_i + \sum\limits_{j=4}^6\psi_{24j}\varepsilon_j\Bigg)\eta_2\eta_4 \nonumber\\
& \quad+ \frac14  \Bigg(\sum\limits_{i=1}^3  \psi_{12i}\varepsilon_i + \sum\limits_{j=4}^6\psi_{12j}\varepsilon_j\Bigg)(\eta_1\eta_2 + \eta_3\eta_4)+ \frac14 \Bigg( \sum\limits_{i=1}^3  \psi_{14i}\varepsilon_i
+\sum\limits_{j=4}^6\psi_{14j}\varepsilon_j\Bigg)(\eta_1\eta_4 + \eta_2\eta_3), \nonumber\\
  \hat H_s &= - \sum\limits_q \left( {\cal H}_1
\frac{\sigma_{q1}}{2} + {\cal H}_2 \frac{\sigma_{q2}}{2} + {\cal
H}_3 \frac{\sigma_{q3}}{2} + {\cal H}_4 \frac{\sigma_{q4}}{2}
\right). \label{Hs} 
\end{align}
In  (\ref{Hs}), the following notations are used:
\bea && {\cal H}_1 = \frac{1}{2}J_{11}\eta_1 + \frac{1}{2}J_{12}\eta_2 +
\frac{1}{2}J_{13}\eta_3 + \frac{1}{2}J_{14}\eta_4 +\mu_{13}^{x}E_1 + \mu_{13}^{y}E_2 + \mu_{13}^{z}E_3\,, \label{H1234}\nonumber\\
&& {\cal H}_2 = \frac{1}{2}J_{22}\eta_2 + \frac{1}{2}J_{12}\eta_1 +
\frac{1}{2}J_{24}\eta_4 + \frac{1}{2}J_{14}\eta_3-  \mu_{24}^{x}E_1 - \mu_{24}^{y}E_2 + \mu_{24}^{z}E_3\,, \nonumber\\
&& {\cal H}_3 = \frac{1}{2}J_{11}\eta_3 + \frac{1}{2}J_{12}\eta_4 +
\frac{1}{2}J_{13}\eta_1+ \frac{1}{2}J_{14}\eta_2-  \mu_{13}^{x}E_1 + \mu_{13}^{y}E_2 - \mu_{13}^{z}E_3\,, \nonumber\\
&& {\cal H}_4 = \frac{1}{2}J_{22}\eta_4 + \frac{1}{2}J_{12}\eta_3 +
\frac{1}{2}J_{24}\eta_2 + \frac{1}{2}J_{14}\eta_1+ \mu_{24}^{x}E_1 - \mu_{24}^{y}E_2 - \mu_{24}^{z}E_3\,. 
\eea
In (\ref{H1234}), $\mu_{13}^{x,y,z}=\mu_{1}^{x,y,z}=\mu_{3}^{x,y,z}$, $\mu_{24}^{x,y,z}=\mu_{2}^{x,y,z}=\mu_{4}^{x,y,z}$  are the effective dipole moments per one pseudospin.
The two-particle cluster approximation is used for calculation of the thermodynamic and dielectric characteristics of GPI. In this approximation, thermodynamic potential is given by:
\bea
&& G = N U_{\text{seed}} + NH^0 + N v \sum\limits_{i=1}^3 \sigma_i \varepsilon_i - k_{\text B} T  \sum\limits_q \bigg[ 2\ln \Sp \re^{-\beta \hat H^{(2)}_{q}}
  - \sum\limits_{f=1}^4\ln \Sp  \re^{-\beta \hat H^{(1)}_{qf}} \bigg],  \label{G}
\eea
where $\hat H^{(2)}_{q}$, $\hat H^{(1)}_{qf}$ are two-particle and one-particle Hamiltonians:
\bea
&& \hat H^{(2)}_{q} = - 2w \left( \frac{\sigma_{q1}}{2} \frac{\sigma_{q2}}{2} + \frac{\sigma_{q3}}{2}\frac{\sigma_{q4}}{2}\right)  - \frac{y_1}{\beta}  \frac{\sigma_{q1}}{2} - \frac{y_2}{\beta} \frac{\sigma_{q2}}{2}  -
\frac{y_3}{\beta}  \frac{\sigma_{q1}}{2} - \frac{y_4}{\beta} \frac{\sigma_{q4}}{2}\,,  \label{H2}\\
&& \hat H^{(1)}_{qf} = - \frac{\bar y_f}{\beta}\frac{\sigma_{qf}}{2}\,, \label{H1}
\eea
where such notations are used:
\bea
&& \hspace{-4ex} y_f = \beta (  \Delta_1 + {\cal H}_f),  \qquad \bar y_f =  \beta \Delta_f + y_f.  \label{yf}
\eea
Here, $\Delta_f$ are the effective cluster fields created by the neighboring bonds
from outside  the cluster. In the cluster approximation, the fields $\Delta_f$ can be determined from the self-consistency
condition:  the mean values of the pseudospins $\langle \sigma_{qf} \rangle$ calculated with the two-particle
and one-particle Gibbs distribution, respectively, should coincide:
\bea
\frac{\Sp  \sigma_{qf} \re^{-\beta \hat H^{(2)}_{q}}}{\Sp  \re^{-\beta \hat H^{(2)}_{q}}} =
\frac{\Sp  \sigma_{qf} \re^{-\beta \hat H^{(1)}_{qf}}}{\Sp  \re^{-\beta \hat H^{(1)}_{qf}}}\,.\label{Sp}
\eea
Hence, based on  (\ref{Sp}), with taking into account (\ref{H2}) and (\ref{H1}), we obtain
\begin{align}
 \eta_{1,3}  & =   \frac{1}{D}(\sinh n_{1} \pm\sinh  n_{2} + a^{2}\sinh n_{3} \pm a^{2}\sinh  n_{4}+a\sinh  n_{5} +a \sinh  n_{6}\mp  a\sinh  n_{7} \pm  a\sinh  n_{8})   \nonumber\\
&   =  \tanh  \frac{\bar y_{1,3}}{2}\,, \nonumber \\
 \eta_{2,4}   &=   \frac{1}{D}(\sinh  n_{1} \pm \sinh n_{2} -  a^{2}\sinh  n_{3} \mp  a^{2}\sinh  n_{4}\mp a\sinh  n_{5} \pm  a\sinh  n_{6} +a\sinh  n_{7} +  a\sinh  n_{8})  \nonumber\\
&   =  \tanh  \frac{\bar y_{2,4}}{2}\,,
\label{eta}
\end{align}
where
\begin{align}
D &= \cosh  n_{1} + \cosh  n_{2} + a^{2}\cosh  n_{3} +  a^{2}\cosh  n_{4} +a\cosh  n_{5} + a\cosh  n_{6}
+ a\cosh  n_{7} +  a\cosh  n_{8}\,,\nonumber \\
 a &= \exp\bigg[-\frac{1}{k_{\text B}T}\bigg( w^0 + \sum\limits_{i=1}^3\delta_i\varepsilon_i + \sum\limits_{j=4}^6\delta_j\varepsilon_j \bigg)\bigg], \nonumber\\
 n_{1}&=\frac{1}{2}(y_1 + y_2 + y_3 + y_4 ),\qquad  n_{2}=\frac{1}{2}(y_1 + y_2 - y_3 - y_4), \qquad n_{3}=\frac{1}{2}(y_1 - y_2 + y_3 - y_4), \nonumber\\
 n_{4}&=\frac{1}{2}(y_1 - y_2 - y_3 + y_4),  \qquad  n_{5}=\frac{1}{2}(y_1 - y_2 + y_3 + y_4), \qquad n_{6}=\frac{1}{2}(y_1 + y_2 + y_3 - y_4),\nonumber\\
 n_{7}&=\frac{1}{2}( -y_1 + y_2 + y_3 + y_4), \qquad  n_{8}=\frac{1}{2}(y_1 + y_2 - y_3 + y_4). \nonumber
\end{align}

Taking into consideration  (\ref{yf}), (\ref{eta}), we exclude the parameters $\Delta_f$  and write the relations:
\bea
&& y_1  =  \frac{1}{2} \ln \frac{1  +   \eta_{1}} {1  -   \eta_{1}} + \beta\nu_{11}\eta_{1} + \beta\nu_{12}\eta_{2} + \beta\nu_{13}\eta_{3} + \beta\nu_{14}\eta_{4} +\frac{\beta}{2}(\mu_{13}^{x}E_1 + \mu_{13}^{y}E_2 + \mu_{13}^{z}E_3),\nonumber\\
&& y_2  =  \beta\nu_{12} \eta_{1} + \frac{1}{2} \ln \frac{1  +   \eta_{2}} {1  -   \eta_{2}} + \beta\nu_{22}\eta_{2} + \beta\nu_{14}\eta_{3} + \beta\nu_{24}\eta_{4} +\frac{\beta}{2}(- \mu_{24}^{x}E_1 - \mu_{24}^{y}E_2 +\mu_{24}^{z}E_3),\nonumber\\
&& y_3  =  \beta\nu_{13} \eta_{1} + \beta\nu_{14}\eta_{2} + \frac{1}{2} \ln \frac{1  +   \eta_{3}} {1  -    \eta_{3}} + \beta\nu_{11}\eta_{3} + \beta\nu_{12}\eta_{4} +\frac{\beta}{2}(- \mu_{13}^{x}E_1 + \mu_{13}^{y}E_2 - \mu_{13}^{z}E_3),\nonumber\\
&& y_4  =  \beta\nu_{14} \eta_{1} + \beta\nu_{24}\eta_{2} + \beta\nu_{12}\eta_{3} + \frac{1}{2} \ln \frac{1  +   \eta_{4}} {1  -  \eta_{4}} + \beta\nu_{22}\eta_{4}+\frac{\beta}{2}(\mu_{24}^{x}E_1 - \mu_{24}^{y}E_2 - \mu_{24}^{z}E_3),\nonumber
\eea
where $\nu_{ff'}=\frac{J_{ff'}}{4}$.

At the absence of external electric fields
\bea
\eta_{1}=\eta_{3}=\eta_{13}\,, \quad \eta_{2}=\eta_{4}=\eta_{24}\,,\quad y_{1}=y_{3}=y_{13}\,, \quad y_{2}=y_{4}=y_{24}.\nonumber
\eea

\section{Thermodynamic characteristics of GPI}

To calculate  the dielectric, piezoelectric and elastic characteristics of the GPI we use the thermodynamic potential per one primitive cell obtained in the two-particle cluster approximation:
  \begin{align}
  g  &=  \frac{G}{N}  =  U_{\text{seed}} + H^{0} -  2 \bigg(w^{0}  +  \sum\limits_{i=1}^3 \delta_{i}\varepsilon_i  +   \sum\limits_{j=4}^6 \delta_{j}\varepsilon_i\bigg)- \frac{1}{2}k_{\text B}T \sum\limits_{f=1}^4\ln \bigl( 1 - \eta_{f}^{2}) \nonumber\\
&\quad   - 2k_{\text B}T \ln D + 2k_{\text B}T\ln2 +  v p\sum\limits_{i=1}^3  \varepsilon_i.
\label{GG}
 \end{align}
%
%
From equilibrium conditions, we have obtained equations for the strains $\varepsilon_{i}$, $\varepsilon_{j}$:
 \begin{align}
  -p  &=  c_{l1}^{E0}\varepsilon_1  +  c_{l2}^{E0}\varepsilon_2  +  c_{l3}^{E0}\varepsilon_3  +  c_{l5}^{E0}\varepsilon_5  -  e_{2l}^0E_2  - \frac{2\delta_{l}}{\upsilon} +  \frac{2\delta_l}{v D}M_{\varepsilon}-  \frac{\psi_{11l}}{8v} (\eta_{1}^{2}+\eta_{3}^{2})-\frac{\psi_{13l}}{4v} \eta_{1}\eta_{3}   \nonumber\\
& -\frac{\psi_{22l}}{8v} (\eta_{2}^{2}+\eta_{4}^{2})-   \frac{\psi_{24l}}{4v} \eta_{2}\eta_{4} - \frac{\psi_{12l}}{4v} (\eta_{1}\eta_{2}+\eta_{3}\eta_{4})- \frac{\psi_{14l}}{4v} (\eta_{1}\eta_{4}+\eta_{2}\eta_{3}),\quad (l=1,2,3), \nonumber\\
   0 & =  c_{51}^{E0}\varepsilon_1  +  c_{52}^{E0}\varepsilon_2  +  c_{53}^{E0}\varepsilon_3  +  c_{55}^{E0}\varepsilon_5  -  e_{25}^0E_2  - \frac{2\delta_{5}}{\upsilon} +  \frac{2\delta_5}{v D}M_{\varepsilon}-  \frac{\psi_{115}}{8v} (\eta_{1}^{2}+\eta_{3}^{2})-\frac{\psi_{135}}{4v} \eta_{1}\eta_{3}  \nonumber \\
& -\frac{\psi_{225}}{8v} (\eta_{2}^{2}+\eta_{4}^{2})-   \frac{\psi_{245}}{4v} \eta_{2}\eta_{4} - \frac{\psi_{125}}{4v} (\eta_{1}\eta_{2}+\eta_{3}\eta_{4})- \frac{\psi_{145}}{4v} (\eta_{1}\eta_{4}+\eta_{2}\eta_{3}), \nonumber\\
   0 &= c_{44}^{E0}\varepsilon_4+c_{46}^{E0}\varepsilon_6
 - e_{14}^0 E_1 - e_{34}^0 E_3-\frac{2\delta_{4}}{\upsilon}+ \frac{2\delta_4}{v D}M_{\varepsilon} -  \frac{\psi_{114}}{8v} (\eta_{1}^{2}+\eta_{3}^{2})-  \frac{\psi_{134}}{4v} \eta_{1}\eta_{3} \nonumber\\
& -\frac{\psi_{224}}{8v} (\eta_{2}^{2}+\eta_{4}^{2})-  \frac{\psi_{244}}{4v} \eta_{2}\eta_{4} - \frac{\psi_{124}}{4v} (\eta_{1}\eta_{2}+\eta_{3}\eta_{4})- \frac{\psi_{144}}{4v} (\eta_{1}\eta_{4}+\eta_{2}\eta_{3}),\nonumber\\
   0 &= c_{46}^{E0}\varepsilon_4 +
c_{66}^{E0}\varepsilon_6 - e_{16}^0 E_1 - e_{36}^0 E_3 -\frac{2\delta_{6}}{\upsilon}+ \frac{2\delta_6}{v D}M_{\varepsilon}-  \frac{\psi_{116}}{8v} (\eta_{1}^{2}+\eta_{3}^{2})-\frac{\psi_{136}}{4v} \eta_{1}\eta_{3} \nonumber\\
& -\frac{\psi_{226}}{8v} (\eta_{2}^{2}+\eta_{4}^{2})-    \frac{\psi_{246}}{4v} \eta_{2}\eta_{4} - \frac{\psi_{126}}{4v} (\eta_{1}\eta_{2}+\eta_{3}\eta_{4})- \frac{\psi_{146}}{4v} (\eta_{1}\eta_{4}+\eta_{2}\eta_{3}),\label{sigma}
\end{align}
here, such a notation is used
\bea
&& M_{\varepsilon}  =  2a^{2}\cosh n_{3}+ 2a^{2}\cosh n_{4}
+a \cosh n_{5}+ a\cosh n_{6}+a \cosh n_{7} + a\cosh n_{8}.\nonumber
 \eea
 Differentiating (\ref{GG}) over the fields $E_{i}$, we get the expressions for polarizations $P_{i}$:
\bea
&& P_1   =  e_{14}^0\varepsilon_4   +  e_{16}^0\varepsilon_6  + \chi_{11}^{\varepsilon 0}E_1 +  \chi_{31}^{\varepsilon 0}E_3 +  \frac{1}{2v}[\mu_{13}^{x}(\eta_{1} - \eta_{3}) -  \mu_{24}^{x}(\eta_{2}  - \eta_{4})], \non
 && P_2  =  e_{21}^0\varepsilon_1 + e_{22}^0\varepsilon_2 +
 e_{23}^0\varepsilon_3 + e_{25}^0\varepsilon_5  + \chi_{22}^{\varepsilon 0}E_2  + \frac{1}{2v}[\mu_{13}^{y}(\eta_{1}+\eta_{3})-\mu_{24}^{y}(\eta_{2}+\eta_{4})],\nonumber\\
 && P_3  =    e_{34}^0\varepsilon_4  +  e_{66}^0\varepsilon_6 +   \chi_{33}^{\varepsilon 0}E_3  +  \chi_{31}^{\varepsilon 0}E_1 +  \frac{1}{2v}[\mu_{13}^{z}(\eta_{1} - \eta_{3}) + \mu_{24}^{z}(\eta_{2} - \eta_{4})].\label{P}
 \eea
Static isothermic dielectric susceptibilities of the mechanically clamped crystal
GPI are given by:
\begin{align}
\chi_{11,33}^{\varepsilon}&=\chi_{11, 33}^{\varepsilon 0}+\frac{\beta}{2\upsilon\Delta_{1,3}}\big\{(\mu_{13}^{x,z})^{2}
[\widetilde{D}\lambda_{24}-(\lambda_{13}\lambda_{24}-\lambda^{2})\varphi_{24}^{-}]+(\mu_{24}^{x,z})^{2}[\widetilde{D}\lambda_{13}-(\lambda_{13}\lambda_{24}-\lambda^{2})\varphi_{13}^{-}]\nonumber\\
&\mp2\mu_{13}^{x,z}\mu_{24}^{x,z}[\widetilde{D}\lambda+(\lambda_{13}\lambda_{24}-\lambda^{2})\beta\nu_{2}^{-}]\big\},\\
\Delta_{1,3}&=\widetilde{D}^{2}-\widetilde{D}(\lambda_{24}\varphi_{13}^{-}+\lambda_{13}\varphi_{24}^{-}+2\lambda\beta\nu_{2}^{-})+(\lambda_{13}\lambda_{24}-\lambda^{2})[\varphi_{13}^{-}\varphi_{24}^{-}-(\beta\nu_{2}^{-})^{2}],\nonumber\\
\chi_{22}^{\varepsilon}&=\chi_{22}^{\varepsilon 0}+\frac{\beta}{2\upsilon\Delta_{2}}\big\{(\mu_{13}^{y})^{2}
[\widetilde{D}\varkappa_{13}-(\varkappa_{13}\varkappa_{24}-\varkappa^{2})\varphi_{24}^{+}]+(\mu_{24}^{y})^{2}[\widetilde{D}\varkappa_{24}-(\varkappa_{13}\varkappa_{24}-\varkappa^{2})\varphi_{13}^{+}]\nonumber\\
&-2\mu_{13}^{y}\mu_{24}^{y}[\widetilde{D}\varkappa+(\varkappa_{13}\varkappa_{24}-\varkappa^{2})\beta\nu_{2}^{+}]\big\},\\
\Delta_{2}&=\widetilde{D}^{2}-\widetilde{D}(\varkappa_{13}\varphi_{13}^{+}+\varkappa_{24}\varphi_{24}^{+}+2\varkappa\beta\nu_{2}^{+})+(\varkappa_{13}\varkappa_{24}-\varkappa^{2})[\varphi_{13}^{+}\varphi_{24}^{+}-(\beta\nu_{2}^{+})^{2}].\nonumber
\end{align}
Here, the following notations are used: 
\bea
&&\widetilde{D}=\cosh(y_{13}+y_{24})+a^{2}\cosh(y_{13}-y_{24})+
2a\cosh y_{13}+2a\cosh y_{24}+ a^{2}+1,\nonumber \\
&& \varphi_{13}^{\pm}=\frac{1}{1 -  \eta_{13}^{2}} +\beta\nu_{1}^{\pm},\quad \varphi_{24}^{\pm}=\frac{1}{1 -  \eta_{24}^{2}} +\beta\nu_{3}^{\pm}; \nonumber\\
&&\nu_{l}^{\pm} = \nu_{l}^{0\pm} +  \bigg(\sum\limits_{i=1}^3\psi_{li}^{\pm}\varepsilon_{i} \pm
\sum\limits_{j=4}^6\psi_{lj}^{\pm}\varepsilon_{j}\bigg),~(l=1,2,3),~~~~ \nu_{1}^{0\pm} = \frac{1}{4}(J_{11}^{0} \pm  J_{13}^{0});~~~~\psi_{1i}^{\pm}=\frac{1}{4}(\psi_{11i} \pm \psi_{13i}),\nonumber\\
&&\nu_{2}^{0\pm}=\frac{1}{4}(J_{12}^{0}\pm J_{14}^{0});~~~~\psi_{2i}^{\pm}=\frac{1}{4}(\psi_{12i}\pm\psi_{14i}); ~~~~\nu_{3}^{0\pm}=\frac{1}{4}(J_{22}^{0}\pm J_{24}^{0});~~~~\psi_{3i}^{\pm}=\frac{1}{4}(\psi_{22i}\pm\psi_{24i}),\nonumber\\
&&\lambda_{13}=1+a^{2}+2a\cosh y_{13}\,, ~~~~ \lambda_{24}=1+a^{2}+2a\cosh y_{24}\,,~~~~\lambda=1-a^{2},\nonumber\\
&&\varkappa_{13}=\cosh  (y_{13}+y_{24})+a^{2}\cosh  (y_{13}-y_{24})+
2a\cosh  y_{13}- \eta_{13}^{2}\widetilde{D},\nonumber\\
&&\varkappa_{24}=\cosh  (y_{13}+y_{24})+a^{2}\cosh  (y_{13}-y_{24})+
2a\cosh  y_{24}- \eta_{24}^{2}\widetilde{D},\nonumber\\
&&\varkappa=\cosh  (y_{13}+y_{24})-a^{2}\cosh  (y_{13}-y_{24})
- \eta_{13}\eta_{24}\widetilde{D}.\nonumber
\eea
Based on  (\ref{P}), we have obtained  expressions for isothermic coefficients of piezoelectric stress  $e_{2l}$ ($l=1,2,3,5$) of GPI:
\bea
 && e_{2l} =   \left( \frac{\partial P_2}{\partial
 \varepsilon_l}\right)_{E_2}   = e_{2l}^0+ \frac{\mu_{13}^{y}}{v}\frac{\beta}{\Delta_{2}}\big[(\psi_{1l}\eta_{13} + \psi_{2l}\eta_{24})\tau_{1}^{\psi}
 + (\psi_{2l}\eta_{13} + \psi_{3l}\eta_{24})\tau_{2}^{\psi} - 2\delta_{l}\tau_{1}^{\delta}\big]\nonumber\\
&& ~~~~~-\frac{\mu_{24}^{y}}{v}\frac{\beta}{\Delta_{2}}\big[(\psi_{1l}\eta_{13} + \psi_{2l}\eta_{24})\tau_{2}^{\psi}
 + (\psi_{2l}\eta_{13} + \psi_{3l}\eta_{24})\tau_{3}^{\psi} - 2\delta_{l}\tau_{2}^{\delta}\big],
\eea
where
\bea
&&\tau_{1}^{\psi}  = \widetilde{D}\varkappa_{13}  - (\varkappa_{13}\varkappa_{24}  - \varkappa^{2})\varphi_{24}^{+}\,,
 ~~~~ \tau_{2}^{\psi} = \widetilde{D}\varkappa + (\varkappa_{13}\varkappa_{24} - \varkappa^{2})\beta\nu_{2}^{+}, ~~~~
 \tau_{3}^{\psi} = \widetilde{D}\varkappa_{24} - (\varkappa_{13}\varkappa_{24} - \varkappa^{2})\varphi_{13}^{+}\,,\nonumber\\
&& \tau_{1}^{\delta} =[\widetilde{D}-(\varkappa_{24}\varphi_{24}^{+} +\varkappa\beta\nu_{2}^{+})]\rho_{13}+
(\varkappa\varphi_{24}^{+}+\varkappa_{13}\beta\nu_{2}^{+})\rho_{24}\,,\nonumber\\
&& \tau_{2}^{\delta}=[\widetilde{D}-(\varkappa_{13}\varphi_{13}^{+}+\varkappa\beta\nu_{2}^{+})]\rho_{24}
+(\varkappa\varphi_{13}^{+} +\varkappa_{24}\beta\nu_{2}^{+})\rho_{24}\,,\nonumber\\
&&\rho_{13}=[a^{2}\sinh (y_{13}-y_{24})+a\sinh y_{13}]-\eta_{13}M,~ \nonumber\\
&&\rho_{24}=[-a^{2}\sinh (y_{13}-y_{24})+a\sinh y_{24}]-\eta_{24}M,\nonumber\\
 && M=a^{2}\cosh (y_{13}-y_{24})+a \cosh y_{13}+a\cosh y_{24}+a^{2}.\nonumber
 \eea
Proton contribution to elastic constants of GPI is found by differentiating  (\ref{sigma}) over strains at a constant field:
\begin{align}
 c_{ij}^E &= \left( \frac{\partial \sigma_i}{\partial \varepsilon_i} \right)_{E_2} = c_{ij}^{E0} -\frac{2\beta}{v\Delta_{2}}\big\{(\psi_{1i}\eta_{13}+\psi_{2i}\eta_{24})(\psi_{1j}\eta_{13}+
\psi_{2j}\eta_{24})\tau_{1}^{\psi}\nonumber\\
&+[(\psi_{1i}\eta_{13}+\psi_{2i}\eta_{24})(\psi_{2j}\eta_{13}+
\psi_{3j}\eta_{24})+(\psi_{2i}\eta_{13}+\psi_{3i}\eta_{24})(\psi_{1j}\eta_{13}+
\psi_{2j}\eta_{24})]\tau_{2}^{\psi}\nonumber\\
&+(\psi_{2i}\eta_{13}+\psi_{3i}\eta_{24})(\psi_{2j}\eta_{13}+
\psi_{3j}\eta_{24})\tau_{3}^{\psi}\big\}
+\frac{4\beta\delta_{i}}{\upsilon\Delta_{2}}[(\psi_{1j}\eta_{13} +
\psi_{2j}\eta_{24})\tau_{1}^{\delta} + (\psi_{2j}\eta_{13} +
\psi_{3j}\eta_{24})\tau_{2}^{\delta}] \nonumber\\
&+\frac{4\beta\delta_{j}}{\upsilon\Delta_{2}}[(\psi_{1i}\eta_{13} +
\psi_{2i}\eta_{24})\tau_{1}^{\delta} + (\psi_{2i}\eta_{13} +
\psi_{3i}\eta_{24})\tau_{2}^{\delta}]\nonumber\\
&-\frac{8\beta\delta_{i}\delta_{j}{}}{\upsilon\widetilde{D}\Delta_{2}}[(\rho_{13}\varphi_{13}^{+} + \rho_{24}\beta\nu_{2}^{+})
\tau_{1}^{\delta} +   (\rho_{24}\varphi_{24}^{+} + \rho_{13}\beta\nu_{2}^{+})\tau_{2}^{\delta}]              \nonumber\\
&- \frac{4\beta\delta_{i}\delta_{j}}{\upsilon\widetilde{D}^{2}}\big\{[2a^{2}\cosh (y_{13} - y_{24}) + a \cosh y_{13} + a\cosh y_{24} + 2a^{2}]\widetilde{D}  - 2M^{2} \big\}.
\end{align}
Other dielectric, piezoelectric and elastic characteristics of GPI can be found using the expressions established above. In particular, the matrix of isothermal elastic compliance at a constant field $s_{ij}^E$, which is reciprocal to matrix of elastic constants $c_{ij}^E$:
\bea
&& \widehat{C^E} = \left( \begin{array}{cccc}
            c_{11}^E & c_{12}^E & c_{13}^E & c_{15}^{E}\vspace{0.5ex} \\
            c_{12}^E & c_{22}^E & c_{23}^E & c_{25}^E \vspace{0.5ex}\\
            c_{13}^E & c_{23}^E & c_{33}^E & c_{35}^E \vspace{0.5ex}\\
            c_{15}^E & c_{25}^E & c_{35}^E & c_{55}^E
            \end{array} \right), \qquad
\widehat{S^E} = (\widehat{C^E})^{-1}, \nonumber
\eea
isothermal coefficients of piezoelectric strain
\be
d_{2l} = \sum\limits_{l'} s_{ll'}^E e_{2l'}\,, \quad (l,l'=1, 2, 3, 5),
\ee
isothermal dielectric susceptibility of a mechanically free crystal
\be
\chi_{22}^{\sigma} = \chi_{22}^{\varepsilon} + \sum\limits_{l} e_{2l} d_{2l}\,,
\ee
isothermal constants of piezoelectric strain
\be
h_{2l} = \frac{e_{2l}}{\chi_{22}^{\varepsilon}}\,,
\ee
isothermal constants of piezoelectric strain
 \begin{eqnarray}
 &&          g_{2l} = \frac{d_{2l}}{\chi_{22}^{\sigma}}. ~~
   \end{eqnarray}

Let us consider thermal characteristics of GPI crystal. Molar entropy of the proton subsystem:
\begin{align}
 S   &=  \frac{R}{4}
\bigg[  -  2\ln2  +  \ln\bigl(  1  -  \eta_{13}  \bigr) +  \ln\bigl(  1 -  \eta_{24}  \bigr) + 2\ln\widetilde{D}-  2 (\beta\nu_{1}^{+}\eta_{13} + \beta\nu_{2}^{+}\eta_{24}) \eta_{13}   \nonumber\\
&   -  2 (\beta\nu_{2}^{+}\eta_{13} + \beta\nu_{3}^{+}\eta_{24})  \eta_{24} +   \frac{4w}{T\widetilde{D}} M \bigg],\label{S}
\end{align}
here, \textit{R} is the gas constant.

The molar heat capacity of a proton subsystem of
GPI crystals can be found numerically from the entropy (\ref{S}):
\bea
&&\Delta C^{\sigma} = T \left( \frac{\partial S}{\partial T} \right)_{\sigma}.
\eea

\section{Comparison of the results of numerical calculations with the experimental data}

To calculate  the temperature  dependences of
dielectric and piezoelectric characteristics
of GPI, which are calculated below, we need to set  certain values of the following parameters:
\begin{itemize}
  \item parameter of  short-range interactions $w^{0}$;
  \item parameters of  long-range interactions $\nu_{f}^{0\pm}$ ($f=1,2,3$);
  \item deformational potentials $\delta_{i}$,   $\psi_{fi}^{\pm}$ ($f=1,2,3$; $i=1,\ldots,6$);
  \item effective dipole moments
$\mu_{13}^{x}$; $\mu_{24}^{x}$; $\mu_{13}^{y}$; $\mu_{24}^{y}$; $\mu_{13}^{z}$; $\mu_{24}^{z}$;
  \item ``seed'' dielectric susceptibilities $\chi_{ii}^{\varepsilon 0}$, $\chi_{31}^{\varepsilon 0}$  ($i=1, 2, 3$);
  \item ``seed'' coefficients of piezoelectric stress $e_{2i}^0$, $e_{25}^0$, $e_{14}^0$, $e_{16}^0$, $e_{34}^0$, $e_{36}^0$;
  \item ``seed''  elastic constants $c_{ii'}^{E0}$, $c_{jj}^{E0}$, $c_{i5}^{E0}$, $c_{46}^{E0}$  ($i=1, 2, 3$; $i'=1, 2, 3$; $j=4, 5, 6$).
\end{itemize}

The volume of a primitive cell of GPI is the  $\upsilon_{0.0} = 0.601\cdot 10^{-21}$~cm$^3$, $\upsilon_{0.808} = 0.6114\cdot 10^{-21}$~cm$^3$. 

\begin{figure}[!t]
\begin{center}
\includegraphics[scale=0.74]{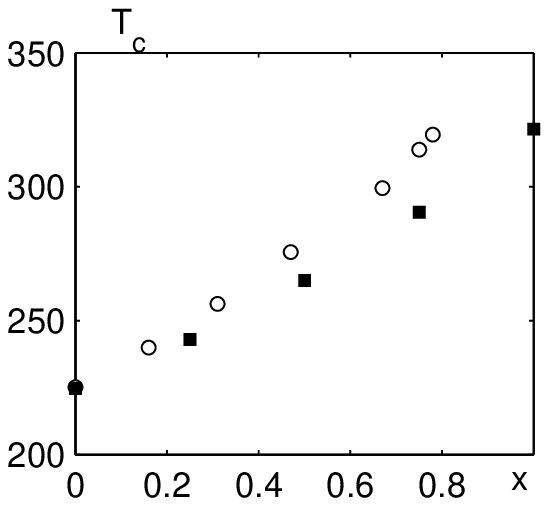}
\end{center}
\caption[]{Phase transition temperature $T_{\text{c}}$ vs deuterium concentration $x$ obtained in GPI$_{1-x}$DGPI$_{x}$: $\circ$~\cite{nay2}; $\blacksquare$ \cite{shi2}.  } \label{Tcx}
\end{figure}

Figure~\ref{Tcx} shows the dependences of the phase transition point $T_{\text{c}}$ of GPI$_{1-x}$DGPI$_{x}$ on the deuteron concentration $x$, which are obtained in {\cite{nay2}} and {\cite{shi2}} and do not agree with each other. The concentration dependence of $T_{\text{c}}$ \cite{nay2} can be approximated by curve $T_{\text{c}}(x) = 225(1+0.382x+0.193x^2)$~K, which is a theoretical curve of $T_{\text{c}}(x)$. Since the papers \cite{nay1,nay2} also present a concentration dependence of spontaneous polarization and dielectric permittivity of the mixed GPI$_{1-x}$DGPI$_{x}$ compounds, in our further analysis  we decided to use $T_{\text{c}}(x)$ published in \cite{nay2}.

In  \cite{yas1}, the temperature $T_{\text{c}}=322.85$~K is stated, that corresponds to the  concentration $x=0.808$ in our model.
The values of the given theory parameters are determined at the study of the static properties of  \cite{Zachek_PB2017}.
The optimal values of  long-range interactions  $\nu_{f}^{0\pm}$ we use $\tilde \nu_1^{0+}=\tilde \nu_2^{0+}=\tilde \nu_3^{0+}=2.643$~K, $\tilde \nu_1^{0-}=\tilde \nu_2^{0-}=\tilde \nu_3^{0-}=0.2$~K,  where $\tilde \nu_{f}^{0\pm}=\nu_{f}^{0\pm}/k_{\text{B}}$. The parameters $\nu_{f}^{0\pm}$ do not depend on concentration $x$.

The calculated parameters $w^{0}$  of the GPI$_{1-x}$DGPI$_{x}$ crystals are  $w_0/k_{\text{B}}=820$~K at $x=0$ and 1323.6~K at $x=0.808$.

The optimal values of the deformational potentials $\delta_{i}$ at $x=0.0$ are
$\tilde\delta_{1}=500$~K,  $\tilde\delta_{2}=600$~K, $\tilde\delta_{3}=500$~K, $\tilde\delta_{4}=150$~K, $\tilde\delta_{5}=100$~K, $\tilde\delta_{6}=150$~K; $\tilde\delta_{i}={\delta_{i}}/{k_{\text{B}}}$.
At $x=0.808$, they are $\tilde\delta_{i}(0.808)=0.337\tilde\delta_{i}(0)$.

The optimal values of the $\psi_{fi}^{\pm}$ are as follows:
$\tilde\psi_{f1}^{+} = 87.9$~K,  $\tilde\psi_{f2}^{+} = 237.0$~K,  $\tilde\psi_{f3}^{+} = 103.8$~K,
$\tilde\psi_{f4}^{+} = 149.1$~K,  $\tilde\psi_{f5}^{+} = 21.3$~K,  $\tilde\psi_{f6}^{+} = 143.8$~K,  $\tilde\psi_{fi}^{-}=0$~K, where
 $\tilde\psi_{fi}^{\pm} =\psi_{fi}^{\pm}/{k_{\text{B}}}$.
At $x=0.808$,  they are  $\psi_{fi}^{\pm}(0.808)=0.337\psi_{fi}^{\pm}(0)$.

The effective dipole moments in the paraelectric phase are equal to ${\bm\mu}_{13} = (0.4,4.02,4.3)\cdot 10^{-18}$~esu$\cdot$cm,  ${\bm\mu_{24}} = (-2.3,-3.0,2.2)\cdot 10^{-18}$~esu$\cdot$cm and do not depend on deuteration. In the ferroelectric phase, the $y$-component of the first dipole moment increases on deuteration as $\mu_{\text{13ferro}}^{y}(x)=3.82(1+0.062x)\linebreak\times10^{-18}$~esu$\cdot$cm, and at $x=0.808$, it is $\mu_{\text{13ferro}}^{y}(x=0.808) =4.01\cdot 10^{-18}$~esu$\cdot$cm.

``Seed'' coefficients of piezoelectric stress, dielectric susceptibilities and elastic constants\\
$e_{21}^0=e_{22}^0=e_{23}^0=e_{25}^0=e_{14}^0=e_{16}^0=e_{34}^0=e_{36}^0=0.0~\frac{\text{esu}}{\text{cm}^2}$; \\
$\chi_{11}^{\varepsilon 0}=0.1$, ~~~$\chi_{22}^{\varepsilon 0}(x=0.0)=0.403$, ~~~$\chi_{22}^{\varepsilon 0}(x=0.808)=2.2$, ~~~$\chi_{33}^{\varepsilon 0}=0.5$,~~~ $\chi_{31}^{\varepsilon 0}=0.0$;\\
$c_{11}^{0E}   =   26.91\cdot10^{10}~\frac{\text{dyn}}{\text{cm}^2}$\,,~~~
$c_{12}^{E0}   =   14.5 \cdot 10^{10}~\frac{\text{dyn}}{\text{cm}^2}$\,,~~~
$c_{13}^{E0}   =   11.64 \cdot10^{10}~\frac{\text{dyn}}{\text{cm}^2}$\,,~~~
$c_{15}^{E0} = 3.91  \cdot10^{10}~\frac{\text{dyn}}{\text{cm}^2}$\,,\\
$c_{22}^{E0} = [64.99 -  0.04(T-T_{\text{c}})] \cdot10^{10}~\frac{\text{dyn}}{\text{cm}^2}$\,,~~~
$c_{23}^{E0} = 20.38\cdot10^{10}~ \frac{\text{dyn}}{\text{cm}^2}$\,,~~~
$c_{25}^{E0} = 5.64  \cdot10^{10}~\frac{\text{dyn}}{\text{cm}^2}$\,, \\
$c_{33}^{E0} = 24.41\cdot10^{10} ~\frac{\text{dyn}}{\text{cm}^2}$\,,~~~
$c_{35}^{E0} = -2.84  \cdot10^{10}~\frac{\text{dyn}}{\text{cm}^2}$\,,~~~
$c_{55}^{E0} = 8.54 \cdot 10^{10}~\frac{\text{dyn}}{\text{cm}^2}$\,,\\
$c_{44}^{E0} = 15.31 \cdot10^{10}~\frac{\text{dyn}}{\text{cm}^2}$\,,~~~
$c_{46}^{E0} = -1.1 \cdot 10^{10}~\frac{\text{dyn}}{\text{cm}^2}$\,,~~~
$c_{66}^{E0} = 11.88 \cdot 10^{10}~\frac{\text{dyn}}{\text{cm}^2}$\,.

Now, let us focus on the obtained results and analyse the effect of  hydrostatic pressure $p=-\sigma_1=-\sigma_2=-\sigma_3$ on thermodynamic characteristics of GPI$_{1-x}$DGPI$_{x}$.

\begin{figure}[!t]
\begin{center}
  \includegraphics[scale=0.69]{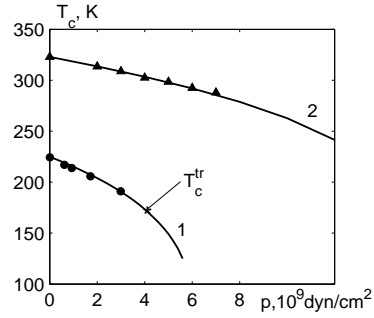}
\end{center}
\caption[]{The pressure dependence of the temperature $T_{\text{c}}$  of GPI$_{1-x}$DGPI$_{x}$ at different  \textit{x}: 0.00~---~1; $\bullet$~\cite{yas}; 0.808~---~2; $\blacktriangle$  \cite{yas1}. } \label{Tc}
\end{figure}

The pressure dependences of temperature $T_{\text{c}}$  of GPI$_{1-x}$DGPI$_{x}$ at  $x=0.0$  and $x=0.808$ are presented in figure~\ref{Tc}.

The calculated  dependences $T_{\text{c}}(p)$ at the established theory parameters quantitatively well describe the experimental data  \cite{yas,yas1}.
Applying a hydrostatic pressure to the crystals decreases their transition temperature $T_{\text{c}}(p)$. The rate of decreasing of transition temperature with an increase of pressure at  $x=0.00$ is $\rd T_{\text{c}}/ \rd p = - 11$~K/kbar \cite{yas} up to the pressure  $p = 3.5 10^{9}$~dyn/cm$^{2}$ and the corresponding temperature $T_{\text{c}} = 180$~K, and at higher pressures $T_{\text{c}}$, decreases nonlinearly;   at  $x=0.808$, the rate of a decrease is $\rd T_{\text{c}}/ \rd p = - 5.0$~K/kbar \cite{yas1}.

\begin{figure}[!t]
\begin{center}
 \includegraphics[scale=0.69]{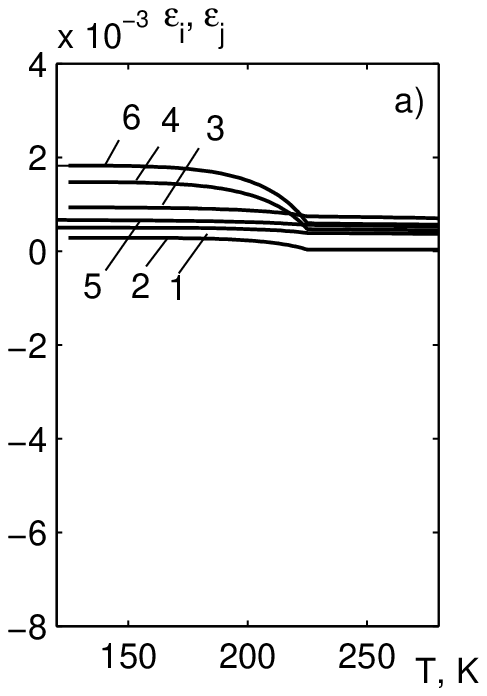}\includegraphics[scale=0.69]{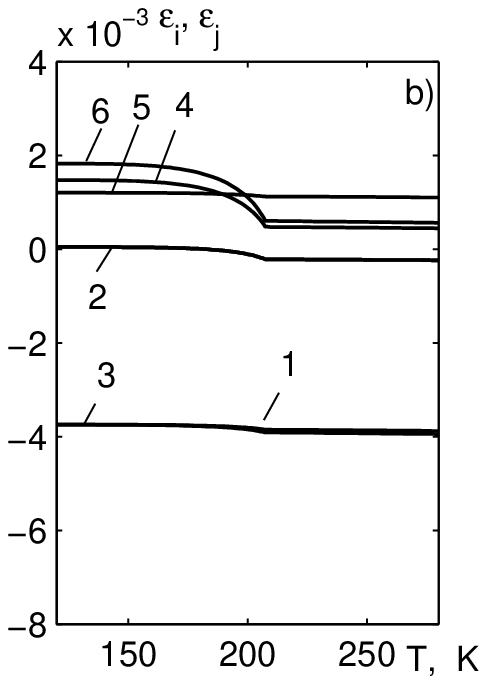}\includegraphics[scale=0.69]{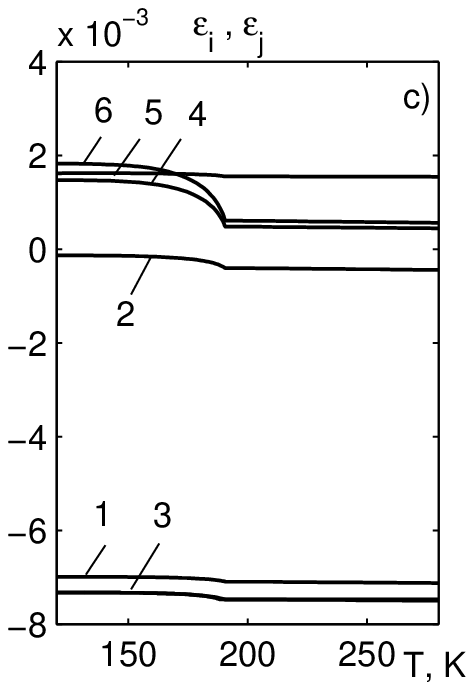}
\end{center}
\caption[]{Temperature dependences of the strains of GPI:
 $\varepsilon_1$~---~1, $\varepsilon_2$~---~2, $\varepsilon_3$~---~3, $\varepsilon_4$~---~4, $\varepsilon_5$~---~5 and $\varepsilon_6$~---~6  at different values of pressure $p$ ($10^{9}$~dyn/cm$^{2}$): 0.0~---~a); 1.7~---~b); 3.0~---~c).} \label{eps}
\end{figure}

Temperature dependences of the strains  $\varepsilon_i$, $\varepsilon_j$ of GPI crystal at different values of hydrostatic pressure~$p$ are presented in figure~\ref{eps}.
The strains $\varepsilon_1$, $\varepsilon_3$ and $\varepsilon_5$ are practically independent of temperature in both phases, but   the strains   $\varepsilon_2$, $\varepsilon_4$ and $\varepsilon_6$ slightly decrease with temperature in the ferroelectric phase and are almost independent of temperature in the paraelectric phase.

\begin{figure}[!t]
\begin{center}
\includegraphics[scale=0.7]{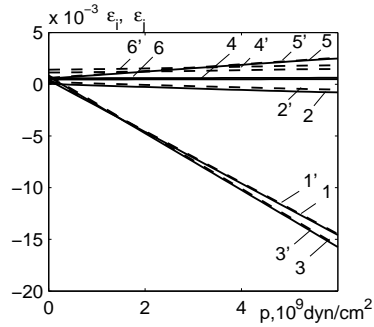}
\end{center}
\vspace{-1mm}
\caption[]{The dependences of the strains of GPI on the pressure at the temperature $T=205$~K:  $\varepsilon_1$~---~1, $\varepsilon_2$~---~2, $\varepsilon_3$~---~3, $\varepsilon_4$~---~4, $\varepsilon_5$~---~5, $\varepsilon_6$~---~6; and  at  $T=245$~K:  $\varepsilon_1$~---~1', $\varepsilon_2$~---~2', $\varepsilon_3$~---~3', $\varepsilon_4$~---~4', $\varepsilon_5$~---~5', $\varepsilon_6$~---~6'.} \label{eiejhh}
\end{figure}

Pressure $p$ leads to a significant increase of absolute values of the strains $\varepsilon_1$ and $\varepsilon_3$, but the other strains depend on $p$ very little (figure~\ref{eiejhh}).

\begin{figure}[!t]
\begin{center}
\includegraphics[scale=0.74]{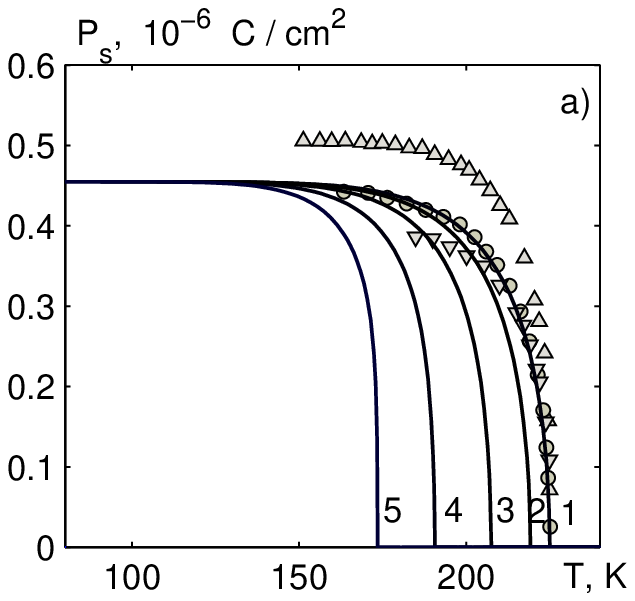}   \includegraphics[scale=0.74]{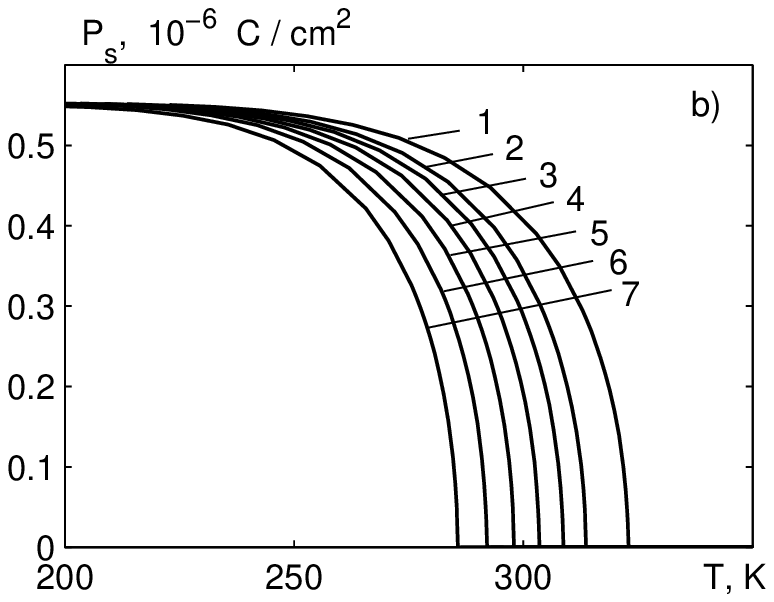}
\end{center}
\vspace{-1mm}
\caption[]{Temperature dependences of the spontaneous polarization of GPI (a) at different values of hydrostatic pressure $p$, ($10^{9}$~dyn/cm$^{2}$):
0.0~---~1, $\circ$ \cite{nay1}, $\vartriangle$ \cite{dac}, $\square$ \cite{wie}, $\triangledown$ \cite{tch1};
0.9~---~2; 1.7~---~3; 3.0~---~4; 4.0~---~5; of GPI$_{0.192}$DGPI$_{0.808}$ (b) at different values of hydrostatic pressure $p$, ($10^{9}$~dyn/cm$^{2}$): 0.0~---~1; 2.0~---~2; 3.0~---~3; 4.0~---~4; 5.0~---~5; 6.0~---~6; 7.0~---~7.}  \label{Ps}
\end{figure}

\begin{figure}[!t]
\begin{center}
 \includegraphics[scale=0.74]{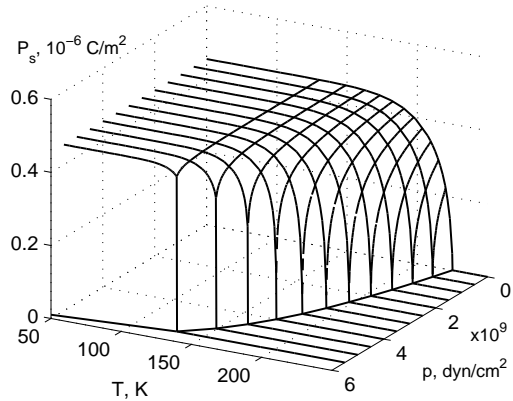}
\end{center}
\vspace{-1mm}
\caption[]{Temperature-pressure dependence of the spontaneous polarization of GPI.} \label{Ptp}
\end{figure}

In figure~\ref{Ps}~(a), the temperature dependences of spontaneous polarization of GPI crystal are presented, in figure~\ref{Ps}~(b) --- for GPI$_{0.192}$DGPI$_{0.808}$ at different values of hydrostatic pressure $p$; in figure~\ref{Ptp} --- the temperature-pressure dependences of  spontaneous polarization of GPI crystal. An increase of $p$ leads to the change of the phase transition order. %

At low pressures, the phase transition is a transition of the second order, but at high pressures, starting with  $p \approx 4$~dyn/cm$^{2}$ (tricritical point in figure~\ref{Tc}), it becomes a transition of the first order. In the case of crystal GPI$_{0.192}$DGPI$_{0.808}$, even at high pressures, there is a second order phase transition.
An increase of $p$ leads to a slight decrease of the  polarization $P_s$ in the whole temperature range.

\begin{figure}[!t]
\begin{center}
\includegraphics[scale=0.74]{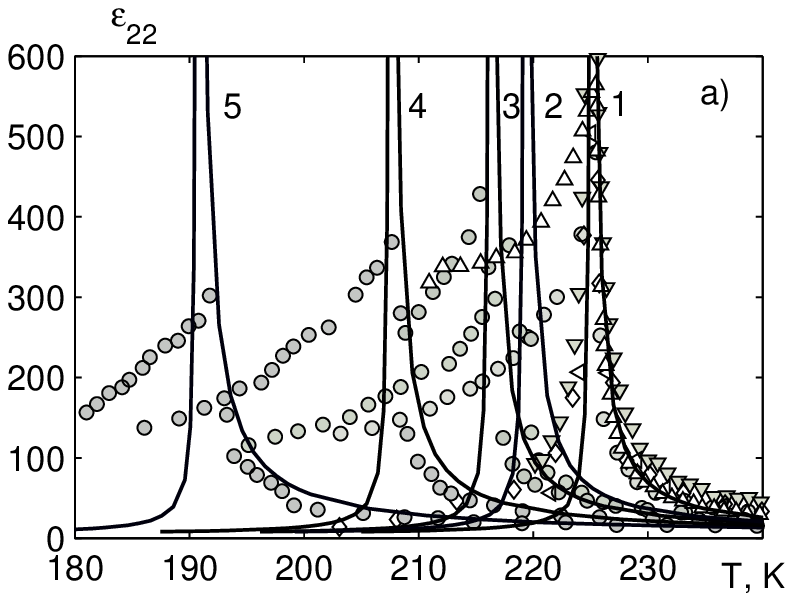} \includegraphics[scale=0.74]{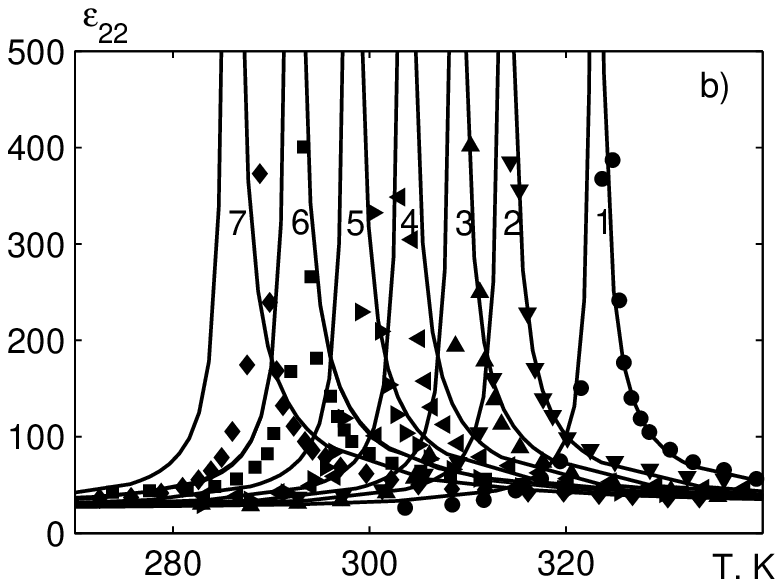}
\end{center}
\vspace{-2mm}
\caption[]{Temperature dependences of the static dielectric permittivity $\varepsilon_{22}^{\varepsilon}$ of GPI crystal (a)  at different values of hydrostatic pressure $p$, ($10^{9}$~dyn/cm$^{2}$): 0.0~---~1, $\lozenge$ \cite{nay2}, $\triangledown$ \cite{tch1}, $\vartriangle$ \cite{dac}, $\square$ \cite{wie}, $\circ$ \cite{yas};
0.6~---~2, $\circ$ \cite{yas}; 0.9~---~3, $\circ$ \cite{yas}; 1.7~---~4,
 $\circ$ \cite{yas}; 3.0~---~5, $\circ$ \cite{yas};  and of GPI$_{0.192}$DGPI$_{0.808}$ (b)  at different values of hydrostatic pressure $p$, ($10^{9}$~dyn/cm$^{2}$): 0.0~---~1, $\bullet$ \cite{yas1}; 2.0~---~2, $\blacktriangledown$ \cite{yas1}; 3.0~---~3, $\blacktriangle$ \cite{yas1}; 4.0~---~4, $\blacktriangleleft$ \cite{yas1}; 5.0~---~5, $\blacktriangleright$ \cite{yas1}; 6.0~---~6, $\blacksquare$ \cite{yas1}; 8.0~---~7 $\blacklozenge$ \cite{yas1}.} \label{eps22p}
\end{figure}

Temperature dependences of the longitudinal static dielectric permittivity of GPI and \linebreak GPI$_{0.192}$DGPI$_{0.808}$ crystals at different values of pressure are presented in figure~\ref{eps22p}~(a) and \ref{eps22p}~(b), respectively. The results of theoretical calculations quantitatively well agree with experimental data \cite{yas,yas1} in the paraelectric phase at small values of hydrostatic pressure $p$. Disagreement in ferroelectric phase for $\varepsilon_{22}^{\varepsilon}$ is connected with domain reorientation contribution to permittivity, which is not taken into account in our theory.

\begin{figure}[!t]
\begin{center}
 \includegraphics[scale=0.69]{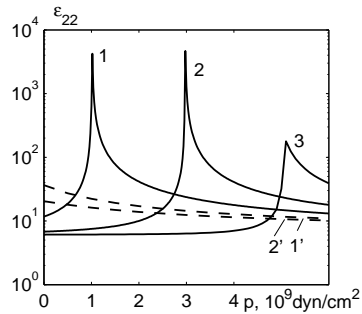}
\end{center}
\vspace{-2mm}
\caption[]{Pressure  dependences of the dielectric permittivity  $\varepsilon_{22}$ of GPI crystal  at different values of temperature  $T$,~K: 245~---~2'; 235~---~1'; 215~---~1; 191~---~2; 159~---~3.} \label{eps22ht}
\end{figure}

The dependences of dielectric permittivity $\varepsilon_{22}$  of GPI crystal on hydrostatic pressure at different values of temperature are presented in figure~\ref{eps22ht}. In the paraelectric phase,  $\varepsilon_{22}$ decreases with an increase of pressure $p$, but in ferroelectric phase,  permittivity $\varepsilon_{22}$ increases up to the phase transition pressure, and then decreases.
\begin{figure}[!t]
\begin{center}
\includegraphics[scale=0.74]{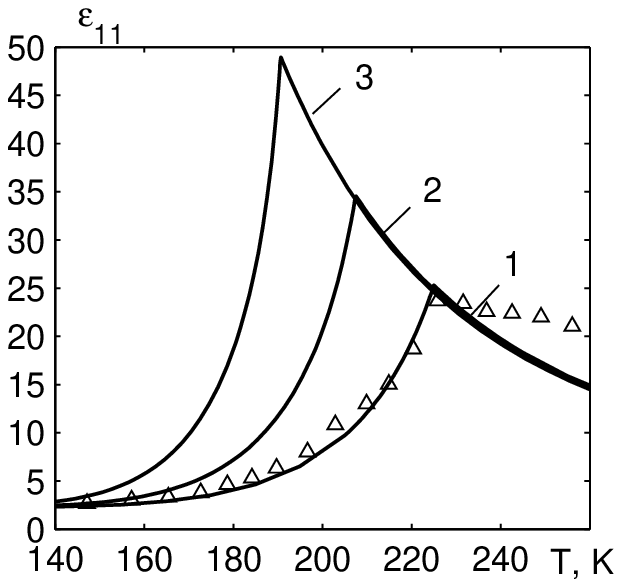}\includegraphics[scale=0.74]{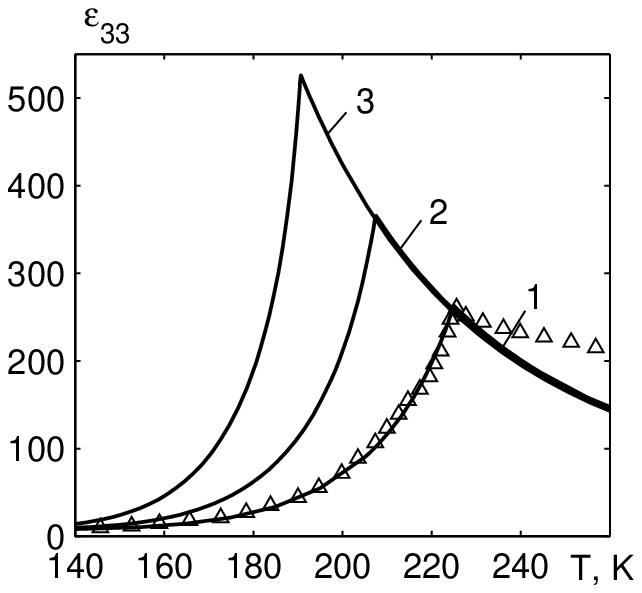} 
\end{center}
\vspace{-2mm}
\caption[]{Temperature dependences of the dielectric permittivities $\varepsilon_{11}$ and $\varepsilon_{33}$  of GPI crystal at different values of hydrostatic pressure $p$, ($10^{9}$~dyn/cm$^{2}$):
0.0~---~1,  $\vartriangle$ \cite{dac};  1.7~---~2; 3.0~---~3.} \label{eps11}
\end{figure}
\begin{figure}[!t]
\begin{center}
\includegraphics[scale=0.74]{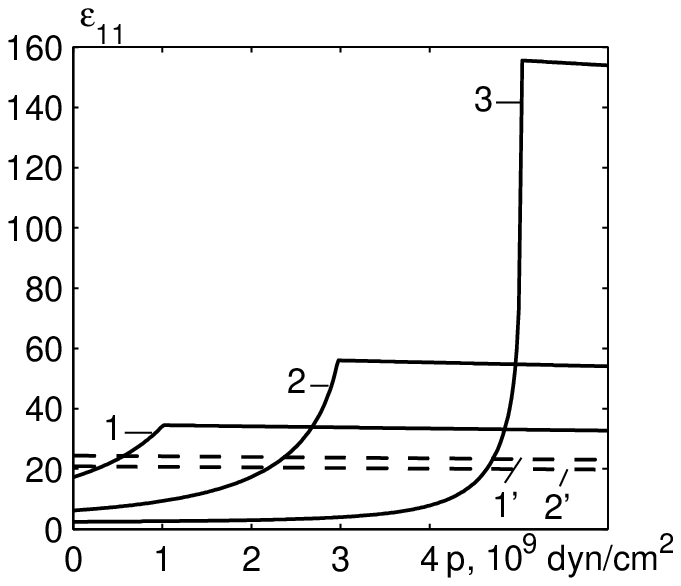} \includegraphics[scale=0.74]{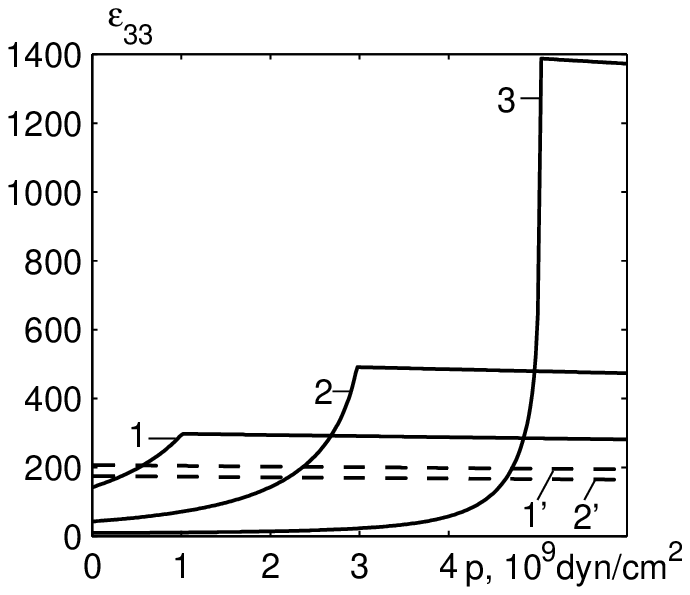}
\end{center}
\vspace{-2mm}
\caption[]{Pressure  dependences of the dielectric permittivities $\varepsilon_{11}$ and $\varepsilon_{33}$  at different values of temperature $T$,~K: 245~---~2'; 235~---~1'; 215~---~1; 205~---~2; 185~---~3.} \label{eps11ht}
\end{figure}
\begin{figure}[!b]
\begin{center}
\includegraphics[scale=0.74]{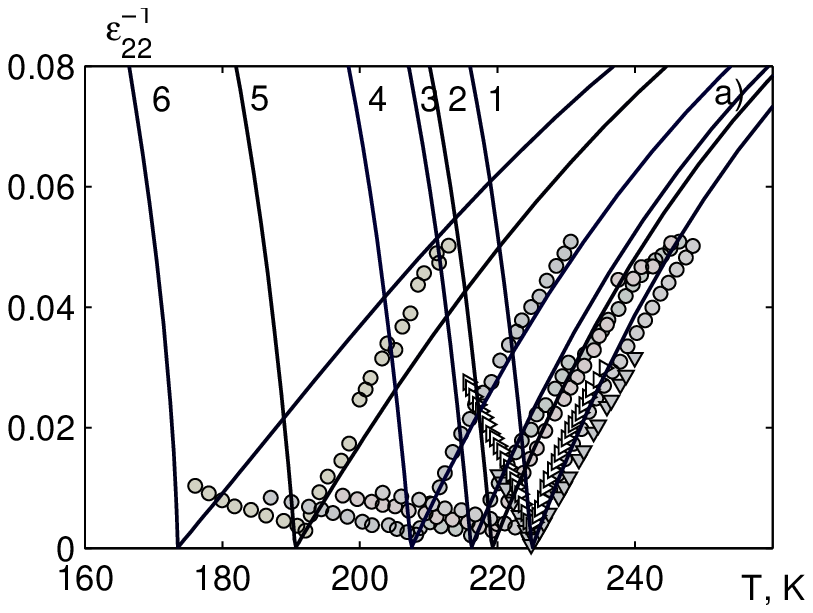}\includegraphics[scale=0.74]{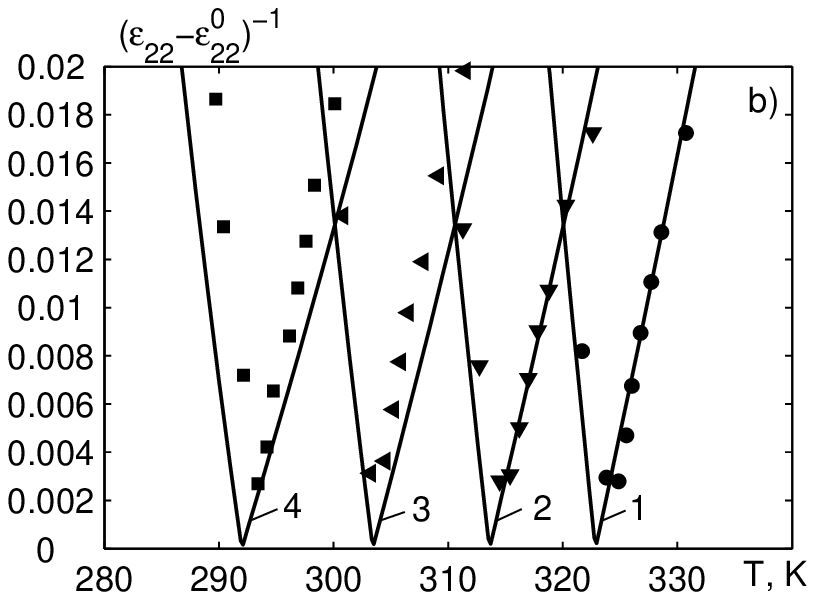} 
\end{center}
\vspace{-3mm}
\caption[]{Temperature dependences of the inverse dielectric permittivity $(\varepsilon_{22}^{\varepsilon})^{-1}$ of GPI crystal (a)  at different values of hydrostatic pressure $p$, ($10^{9}$~dyn/cm$^{2}$): 0.0~---~1, $\triangledown$ \cite{tch1}, $\triangleright$ \cite{Baran1996}, $\circ$ \cite{yas}; 0.6~---~2, $\circ$ \cite{yas}; 0.9~---~3, $\circ$ \cite{yas}; 1.7~---~4, $\circ$ \cite{yas}; 3.0~---~5, $\circ$ \cite{yas}; 4.0~---~6; and GPI$_{0.192}$DGPI$_{0.808}$ (b) at different values of hydrostatic pressure $p$, ($10^{9}$~dyn/cm$^{2}$):  0.0~---~1, $\bullet$ \cite{yas1}, 2.0~---~2, $\blacktriangledown$ \cite{yas1}, 4.0~---~3, $\blacktriangleleft$ \cite{yas1}, 6.0~---~4, $\blacksquare$  \cite{yas1}.} \label{eps22-1}
\end{figure}
\begin{figure}[!b]
\begin{center}
 \includegraphics[scale=0.74]{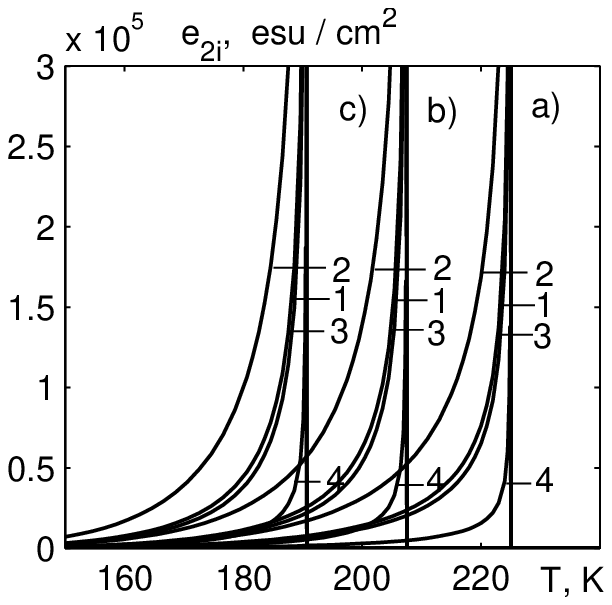} \includegraphics[scale=0.74]{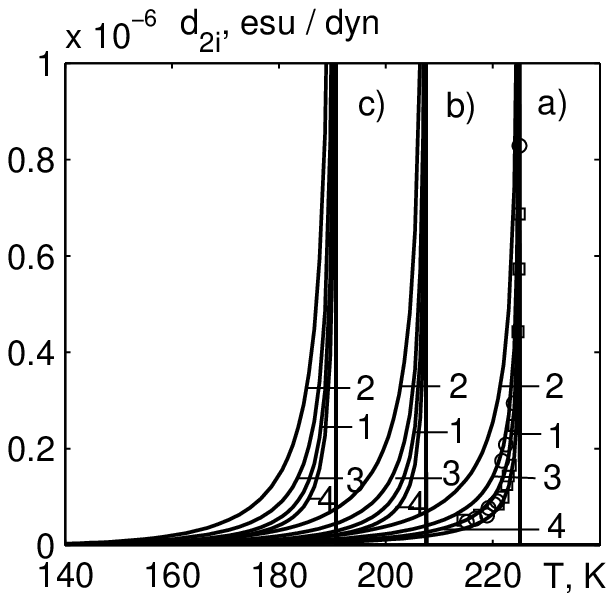}
\end{center}
\vspace{-1mm}
\caption[]{Temperature dependences of the coefficients of piezoelectric stress $e_{2i}$, $e_{25}$ and strain $d_{2i}$, $d_{2i}$: 1~---~$e_{21}$, $d_{21}$ $\square$ \cite{wie}, 2~---~$e_{22}$, $d_{22}$, 3~---~$e_{23}$, $d_{23}$ $\circ$ \cite{wie}, 4~---~$e_{25}$, $d_{25}$ of GPI crystal at  different values of pressure $p$, ($10^{9}$~dyn/cm$^{2}$): 0.0~---~a); 1.7~---~b); 3.0~---~c).} \label{e2i}
\end{figure}

Temperature dependences of  transverse static dielectric permittivities of GPI crystal at different values of hydrostatic pressure are presented in figure~\ref{eps11}, and  pressure  dependences of dielectric permittivity  $\varepsilon_{11}$ and $\varepsilon_{33}$  at different values of temperature  --- in figure~\ref{eps11ht}. 
Notations 1', 2' in figure~\ref{eps11ht} are used for the curves in a paraelectric phase.
The values of $\varepsilon_{11}$ and $\varepsilon_{33}$ increase with an increase of pressure, and maximum values shift to lower  temperatures. In the paraelectric phase, $\varepsilon_{11}$ and $\varepsilon_{33}$ decrease with an increase of pressure $p$, but in ferroelectric phase, transverse permittivities  increase up to the phase transition pressure, and then decrease.

Temperature dependences of the inverse dielectric permittivity  $(\varepsilon_{22}^{\varepsilon})^{-1}$ of GPI and  GPI$_{0.192}$DGPI$_{0.808}$ crystals at different values of pressure $p$ are presented in figure~\ref{eps22-1}~(a) and \ref{eps22-1}~(b), respectively. The results of theoretical calculations quantitatively well agree with experimental data \cite{yas,yas1} in the paraelectric phase at small values of hydrostatic pressure $p$. As was written above, disagreement in ferroelectric phase for $(\varepsilon_{22}^{\varepsilon})^{-1}$ is connected with the domain reorientation contribution to permittivity, which is not taken into account in our theory.

\begin{figure}[!t]
\begin{center}
 \includegraphics[scale=0.74]{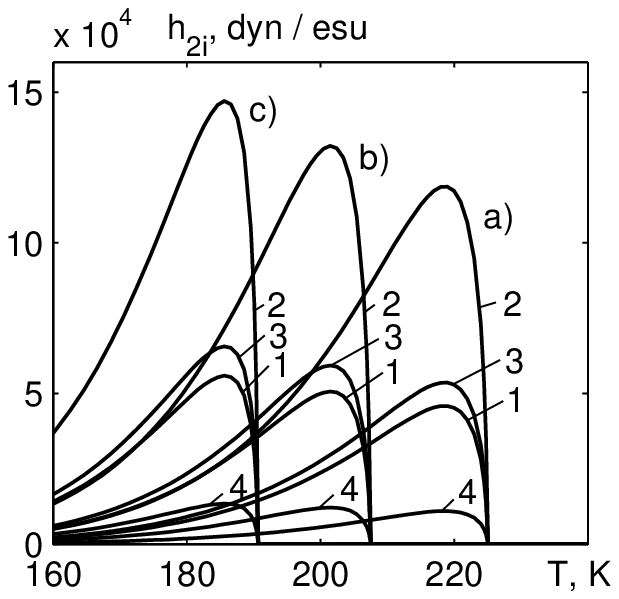}  \includegraphics[scale=0.74]{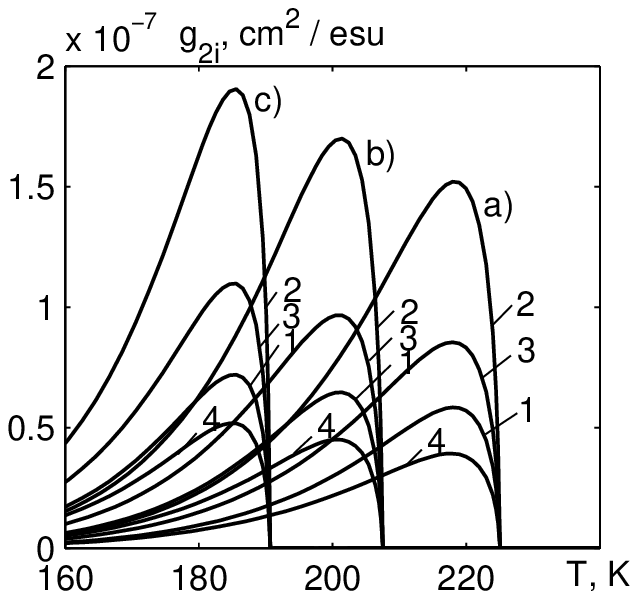}
\end{center}
\vspace{-2mm}
\caption[]{Temperature dependences of the constants of piezoelectric stress $h_{2i}$, $h_{25}$ and strain $g_{2i}$, $g_{25}$: 1~---~$h_{21}$, $g_{21}$, 2~---~$h_{22}$, $g_{22}$, 3~---~$h_{23}$, $g_{23}$, 4~---~$h_{25}$, $g_{25}$ of GPI crystal at  different values of pressure $p$, ($10^{9}$~dyn/cm$^{2}$): 0~---~a); 1.7~---~b); 3~---~c).} \label{h2i}
\end{figure}

Temperature dependences of the coefficients of piezoelectric stress  $e_{2i}$, $e_{25}$ and strain  $d_{2i}$, $d_{2i}$ of GPI crystal at  different values of pressure $p$ are presented in figure~\ref{e2i}; and the temperature dependences of the constants of piezoelectric stress  $h_{2i}$, $h_{25}$ and strain $g_{2i}$, $g_{25}$ --- in figure~\ref{h2i}.

Hydrostatic pressure practically does not influence the magnitude of the $e_{3i}$, $e_{35}$ and $d_{2i}$, $d_{2i}$, but just shifts their maxima to lower temperatures. An increase of the pressure $p$ leads to an increase of magnitude of the piezoelectric coefficients  $h_{2i}$, $h_{25}$ and $g_{2i}$, $g_{25}$.

\begin{figure}[!t]
\begin{center}
 \includegraphics[scale=0.74]{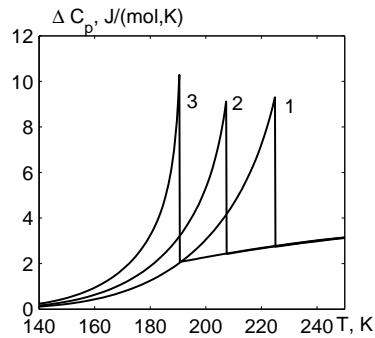}
\end{center}
\vspace{-2mm}
\caption[]{Temperature  dependences of $\Delta C_{p}$
at  different values of hydrostatic pressure $p$, ($10^{9}$~dyn/cm$^{2}$): 0.0~---~1; 0.9~---~2; 1.7~---~3; 3.0~---~4.} \label{Cp}
\end{figure}

Figure~\ref{Cp} shows  temperature  dependences of pseudospin contribution on heat capacity  $\Delta C_{p}$.
In the paraelectric phase, the value of $\Delta C_{p}$ practically does not change with an increase of pressure $p$, but in the ferroelectric phase,  the value of $\Delta C_{p}$ increases with pressure.

\vspace{-2mm}
\section{Conclusions}

In this paper, the effect of hydrostatic pressure on phase transition and physical characteristics  of the GPI$_{1-x}$DGPI$_{x}$ crystals is studied in the frames of two-particle cluster approximation  within the modified proton ordering model of  GPI type quasione-dimensional ferroelectrics with hydrogen bonds, which takes into account the piezoelectric coupling with the strains $\varepsilon_i$, $\varepsilon_j$ in the ferroelectric phase.
We have determined how  the strains $\varepsilon_i$, $\varepsilon_j$ are changed under hydrostatic pressure. These changes of the strains lead to a pressure dependence of the parameters of interactions and, consequently, to a pressure dependence of the transition temperature and other characteristics of these crystals.
At low pressures, the phase transition in our model of GPI is a transition of the second order, but at high pressures, starting from some critical pressure, it becomes a transition of the first order in the nondeuterated crystal. In the case of deuterated crystal, even at high pressures, there is the second order phase transition. The pressure effect in the nondeuterated crystal is much stronger than in a deuterated crystal.
A good quantitative description of the observed pressure and temperature dependences of the considered characteristics has been obtained in paraelectric phase at small values of pressure at the proper choice of the model parameters.

\ukrainianpart

\title{Вплив гідростатичного тиску на термодинамічні    характеристики   сегнетоактивних матеріалів типу NH$_3$CH$_2$COOH$\cdot$H$_2$PO$_3$}
\author{ І.Р. Зачек\refaddr{label1}, Р.Р. Левицький\refaddr{label2}, А.С. Вдович\refaddr{label2}}
\addresses{\addr{label1} Національний університет ``Львівська політехніка'', вул. С. Бандери,  12,  79013  Львів, Україна
\addr{label2} Інститут фізики конденсованих систем НАН України, вул. Свєнціцького,  1, 79011  Львів,  Україна
 }


\vspace{-3mm}
\makeukrtitle

\begin{abstract}
\tolerance=3000%
Для дослідження ефектів, що виникають під дією зовнішніх тисків,
використано модифіковану  модель NH$_3$CH$_2$COOH$\cdot$H$_2$PO$_3$ (GPI)
шляхом врахування п'єзоелектричного зв'язку структурних елементів, які впорядковуються, з
деформаціями  $\varepsilon_i$, $\varepsilon_j$. В наближенні двочастинкового кластера  розраховано компоненти  вектора
поляризації та тензора статичної діелектричної проникності механічно
затиснутого і вільного кристалів, їх п'єзоелектричні  та теплові характеристики. Досліджено вплив гідростатичного  тиску  на фазовий перехід та фізичні характеристики кристалу. Отримано добрий кількісний опис експериментальних даних  для цих кристалів.
\keywords сегнетоелектрики, фазовий перехід, діелектрична проникність, п'єзоелектричні коефіцієнти, гідростатичний тиск

\end{abstract}

\end{document}